\documentclass[a4paper]{jpconf}
\usepackage{graphicx}

\def \znbb {0\nu\beta\beta}

\def \meff {\langle m_{\nu} \rangle}

\def \elam {\langle \lambda \rangle}
\def \eeta{\langle \eta \rangle}
\def \exi{\langle \xi \rangle}
\def\gsim{\raise0.3ex\hbox{$\;>$\kern-0.75em\raise-1.1ex\hbox{$\sim\;$}}}
\def\lsim{\raise0.3ex\hbox{$\;<$\kern-0.75em\raise-1.1ex\hbox{$\sim\;$}}}
\def\rpm{R_P \hspace{-1.0em}/\;\:}

\begin{document}
\hfill{IFIC/06-26}

\title{Phenomenology of neutrinoless double beta decay}

\author{Martin Hirsch}

\address{AHEP Group, IFIC/CSIC, 
  Edificio Institutos de Paterna, Apt 22085, E--46071 Valencia, Spain}

\ead{mahirsch@ific.uv.es}

\begin{abstract}
Neutrinoless double beta decay ($0\nu\beta\beta$) violates lepton number 
by two units, a positive observation therefore necessarily implies physics 
beyond the standard model. Here, three possible contributions to 
$0\nu\beta\beta$ decay are briefly reviewed: (a) The mass mechanism and 
its connection to neutrino oscillations; (b) Left-right symmetric models 
and the lower limit on the right-handed $W$ boson mass; and (c) R-parity 
violating supersymmetry. In addition, the recently published ``extended 
black box'' theorem is briefly discussed. Combined with data from 
oscillation experiments this theorem provides proof that the 
$0\nu\beta\beta$ decay amplitude must receive a non-zero contribution 
from the mass mechanism, if neutrinos are indeed Majorana particles.
\end{abstract}

\section{Introduction}

Since the discovery of neutrino oscillations \cite{Fukuda:1998mi} most 
papers on neutrinoless double beta decay ($\znbb$) have exclusively 
concentrated on its implications for Majorana neutrino masses. However, 
as is well-known, {\em any} model beyond the standard model of particle 
physics, which allows for lepton number violation, potentially 
contributes to $\znbb$ decay. Thus, the basic physics of $\znbb$ decay 
can be summarized as: 
\begin{eqnarray}\label{defbb}
\Big[ {T_{1/2}^{\znbb}} \Big]^{-1} = \Big(\sum_i {\langle \epsilon_i\rangle} 
{ {\cal M}_{\epsilon_i}} \Big)^2{F^{\znbb}}.
\end{eqnarray}
The factor $\langle \epsilon_i\rangle$ contains some (unknown, but lepton 
number violating) particle physics parameters. To determine the numerical 
value of ${\langle \epsilon_i\rangle}$ input from both, experiment and 
theoretical nuclear physics, is needed. Experiments limit (or measure) 
$T_{1/2}^{\znbb}$, for a discussion of various different experiments 
see, for example \cite{bbexp}. ${\cal M}_{\epsilon_i}$ in eq. (\ref{defbb}) 
stands for a nuclear structure matrix element. Different particle 
physics contributions to $\znbb$ decay depend on different matrix 
elements. No definite consensus about the  value and, most importantly, 
the error of nuclear matrix elements exist up to now. For a thorough 
discussion see \cite{simkovic}. Finally, $F^{\znbb}$ is a leptonic phase 
space integral, its value can be calculated quite precisely \cite{Doi:1985dx}. 

This talk concentrates exclusively on particle physics aspects of $\znbb$ 
decay. The classic ``black box'' \cite{Schechter:1981bd} theorem and its 
recently published ``extended'' version \cite{hir2006} are briefly discussed, 
before reviewing constraints on left-right symmetric models and supersymmetry 
with R-parity violation derived from a lower limit on the $\znbb$ decay 
half-live. Last but not least, expectations for the mass mechanism of 
$\znbb$ decay in light of neutrino oscillation data are discussed. It is 
curious to note, that combining the ``extended black box'' with oscillation 
data \cite{hir2006} already today demonstrates that there must be a 
non-zero contribution from the mass mechanism to the $\znbb$ decay 
amplitude, if neutrinos are indeed Majorana particles.

\section{$\znbb$ decay and the Black Box}

From the experimental point of view lepton number violation in $\znbb$ 
decay is observed through the appearance of two electrons in the final 
state with {\bf no} missing energy. Many different, possible mechanisms 
have been discussed in the literature. Interestingly, however, one can 
show \cite{Schechter:1981bd} that independent of which contribution to 
$\znbb$ decay is the dominant one, neutrinos are guaranteed to have a 
non-zero Majorana mass, if $\znbb$ decay is observed. The proof of this 
``black box'' theorem \cite{Schechter:1981bd} essentially follows from 
the observation that any effective low-energy $\Delta L\ne 0$ operator 
inducing $\znbb$ decay will contribute also - possibly at the some order in 
perturbation theory, for sure in some higher order - to the ($\nu_e-\nu_e$) 
entry of the Majorana neutrino mass matrix ($M^{\nu}_{ee}$). A perfect 
cancellation of all different contributions to $M^{\nu}_{ee}$ would then 
require a special symmetry and the proof of the black box theorem is 
completed by showing that no such symmetry can exist \cite{Takasugi:1984xr} 
in any gauge model containing the standard model charged current interaction.

This well-known theorem has recently been extended to the case of three 
generations of neutrinos and arbitrary lepton number and lepton flavour 
violating processes \cite{hir2006}. Combined with data from oscillation 
experiments this ``extended'' black box theorem can be used to show 
that $M^{\nu}_{ee} \ne 0$. The proof involves two steps. In the first step 
it is shown that any effective operator generating lepton number violating 
processes of the form $\Phi_k \rightarrow \Phi_m l_{\alpha}l_{\beta}$, 
where $\Phi_k$ and $\Phi_m$ stand symbolically for any set of SM particles 
with $L=0$, necessarily generates a non-zero $M^{\nu}_{\alpha\beta}$ entry 
in the Majorana neutrino mass matrix in higher order of perturbation 
theory. As for the original black box, one can show that there is no 
possible symmetry allowing for a perfect cancellation of different 
contributions to this entry. In the second step, then all allowed neutrino 
mass matrices with $M^{\nu}_{ee}\equiv 0$ are constructed. It is then 
easy to show that none of the possible five structures is consistent 
with oscillation data. One can thus conclude that $M^{\nu}_{ee}\ne 0$ is 
guaranteed for Majorana neutrinos \cite{hir2006} already today. 

The above theorem(s) do not state which mechanism of $\znbb$ decay 
is the dominant one. Two instructive examples, in which the mass 
mechanism might indeed not be the dominant contribution to $\znbb$ decay, 
are therefore discussed next. 

\subsection{Left-right symmetry}

For $\znbb$ decay, with its typical low energy scale of a few MeV, 
all calculations can be done with the effective Hamiltonian 
\cite{Doi:1985dx}
\begin{equation}\label{hlrsym}
{\cal H}_W^{CC} = \frac{G_F}{\sqrt{2}}
                \Big\{ J^{\dagger}_{\mu L} j^{-}_{\mu_L} 
                     + \kappa J^{\dagger}_{\mu R} j^{-}_{\mu_L} 
                  + \eta J^{\dagger}_{\mu L} j^{-}_{\mu_R} 
                + \lambda J^{\dagger}_{\mu R} j^{-}_{\mu_R}\Big\}.
\end{equation}
Here, $J^{\dagger}_{\mu_\alpha} = {\overline u} \gamma_{\mu} d_{\alpha}$ 
and $j^{-}_{\mu_\alpha} = {\overline e} \gamma_{\mu} \nu_{\alpha}$ are 
the hadronic and leptonic charged currents, $L/R$ stands for 
$P_{L/R} = \frac{1}{2}(1 \mp \gamma_5)$. $G_F$ is the Fermi constant, 
$\kappa \simeq \eta \simeq \tan\zeta$, i.e. the mixing angle between 
the $W_L$ and $W_R$ bosons, and $\lambda \simeq (m_{W_L}/m_{W_R})^2$. 

\begin{figure}[h]
\includegraphics[width=8pc]{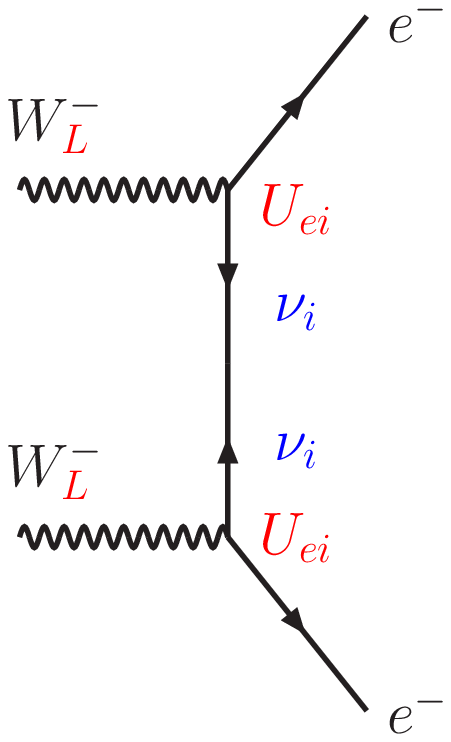}\hspace{1pc}%
\includegraphics[width=8pc]{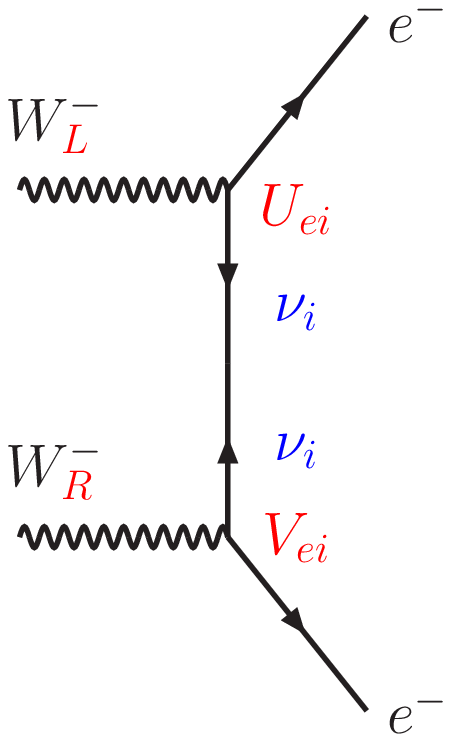}\hspace{1pc}%
\includegraphics[width=8pc]{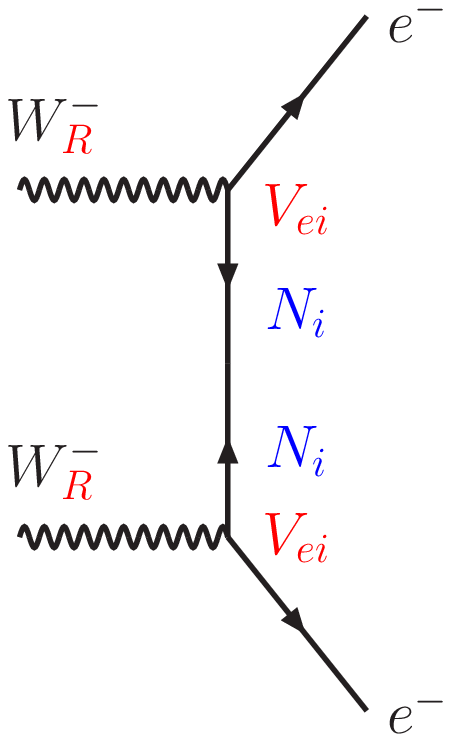}\hspace{1pc}%
\begin{minipage}[b]{10pc}
\caption{\label{fig:lrgraphs}{Leptonic parts of the $\znbb$ decay 
amplitude in left-right symmetric models. The graph to the left 
represents the mass mechansim. The graph in the middle is long-range, 
but suppressed by $\sum_i U_{ei}V_{ei}$. The graph to the right is 
the so-called short range contribution for heavy Majorana neutrinos.}}
\end{minipage}
\end{figure}

The Hamiltonian of eq. (\ref{hlrsym}) gives rise to the diagrams in fig. 
(\ref{fig:lrgraphs}). The graphs on the left and the middle represent 
so-called ``long-range'' contributions. The graph to the left is due to 
a product of two $j^{-}_{\mu_L}$ and corresponds to the mass mechanism 
of $\znbb$ decay, proportional to $\meff= \sum_i U_{ei}^2 m_{\nu_i}$ 
(see discussion in the next section). The graph in the middle is 
proportional to $\elam = \lambda\sum_i U_{ei}V_{ei}$ and $\eeta = 
\eta\sum_i U_{ei}V_{ei}$. The graph to the right is proportional to 
$\exi = \Big[\lambda^2 + \eta^2 - 2 \lambda \eta 
\Big(\frac{M_{GT}^{N}+M_F^N}{M_{GT}^{N}-M_F^N} \Big)\Big]
/\langle m_N\rangle$ \cite{Hirsch:1996qw}.
Here,  $\langle \frac{1}{m_N}\rangle = {\mathop{\sum_j}}'' 
V_{ej}^2 \Big(\frac{m_p}{m_j}\Big)$. 

Formally, the long-range contribution in LR models are suppressed only 
by one power of $\lambda$/$\eta$, compared to the short-range contribution, 
which is quadratic in $\lambda$/$\eta$. Many calculations therefore have 
taken into account only the long-range LR contributions. However, as first 
pointed out by Mohapatra \cite{Mohapatra:1986pj} and confirmed by a 
detailed calculation of the relevant nuclear matrix elements 
\cite{Hirsch:1996qw}, the short-range contribution can be much 
more important then the long-range one. This at first sight contradictive 
statement can be easily understood. In left-right symmetric 
models the mixing between the active, left (and light) neutrinos with 
the heavy, sterile ones can be estimated ``\'a la seesaw'' to be 
very roughly of the order 
$\sum_i U_{ei}V_{ei} \sim \frac{m_D}{M_M} \sim \sqrt{\frac{m_{\nu}}{M_M}}$. 
Then, with a limit of $\elam$ $\lsim 8 \cdot 10^{-7}$ one gets 
$m_{W_R} \gsim 1.1 \hskip1mm m_{W_L} \hskip1mm 
(\frac{m_{\nu}}{\rm 1 eV})^{1/4}(\frac{M_M}{\rm 1 TeV})^{-1/4}$. In 
the short range contribution, although some cancellation of terms 
in $\langle m_N\rangle $ might occur, no such strong supression is 
expected. From \cite{Hirsch:1996qw} and assuming a limit on the $^{76}$Ge 
half-live of $T_{1/2}^{\znbb} \ge 1.2\cdot 10^{25}$ ys a limit of 
\begin{equation}\label{limmwr}
m_{W_R} \gsim 1.3 \Big( \frac{ \langle m_{N}\rangle }
{\rm [1 TeV]}\Big)^{-1/4} \hskip3mm {\rm TeV}
\end{equation}
can then be derived. Note that the limit disappears as $\langle m_N \rangle$ 
goes to infinity, as it should. Note also that the uncertainty in this limit 
due to the uncertainty in the nuclear matrix element calculation scales only 
as $\Delta m_{W_R} \sim (\Delta {\cal M})^{-1/4}$ and thus is quite 
insensitive to the details of the nuclear model.

\subsection{R-parity violation}

In the standard model lepton number is conserved, because there is 
(a) no right-handed neutrino and (b) only one Higgs doublet with 
$L=0$. In supersymmetric models, on the other hand, if one does 
not assume lepton number conservation a priori, one can write 
down the following (trilinear) lepton number violating terms
\begin{eqnarray}\label{Lqqe}
{\cal L}_{\rpm} = &-& \lambda'_{ijk}\left[
        (\bar{u}_L \ \bar{d}_R)_j\cdot
        \mbox{$ \left( \begin{array}{cc}
        e_{R}^{c}\\
        -\nu_{R}^{c}
        \end{array} \right)_i $}\ (\tilde{d}_R)_k
        +
        (\bar{e}_L\ \bar{\nu}_L)_i\ (d_R)_k\cdot
        \mbox{$ \left( \begin{array}{cc}
        \tilde{u}_{L}^{\ast}\\
        -\tilde{d}_{L}^{\ast}
        \end{array} \right)_j $} + \right. \\ \nonumber
        &+& \left. (\bar{u}_L\ \bar{d}_L)_j\ (d_R)_k \cdot
        \mbox{$ \left( \begin{array}{cc}
        \tilde{e}_{L}^{\ast}\\
        -\tilde{\nu}_{L}^{\ast}
        \end{array} \right)_i $}
         + h.c. \right]
\end{eqnarray}
Here, the tilde indicates the scalar superpartners of the usual 
quarks and leptons. A product of two of the terms in eq. (\ref{Lqqe}), 
together with an MSSM neutralino and/or gluino interaction lead 
to $\znbb$ decay diagrams without {\em any} virtual neutrinos being 
exchanged, as first pointed out in \cite{Mohapatra:1986su,Vergados:1986td}. 
A dedicated calculation of all diagrams \cite{Hirsch:1995ek}, together 
with a limit of $T_{1/2}^{\znbb}\ge 1.2 \cdot 10^{25}$ y for 
$^{76}$Ge leads to 
\begin{equation}\label{limlamp}
\lambda'_{111} \le 3.2 \times 10^{-4} (\frac{m_{\tilde q}}{\rm 100 GeV})^2 
(\frac{m_{g}}{\rm 100 GeV})^{1/2}.
\end{equation}
It is interesting to note, that such a small value of $\lambda'_{111}$ 
generates at 1-loop level an entry in the Majorana neutrino mass 
matrix of $M^{\nu}_{ee}\simeq 10^{-6}$ eV only.

\section{Neutrino oscillations and $\znbb$  decay}

If the mass mechanism is dominant, the $\znbb$ decay half-live is 
proportional to the (square of the) ($\nu_e-\nu_e$) element of the 
Majorana neutrino mass matrix. For three generations of light neutrinos, 
this so-called ``effective Majorana'' mass can be expressed as:
\begin{equation}\label{defmeff}
M^{\nu}_{ee}\equiv  \meff = 
      c_{12}^2 c_{13}^2 m_1 + s_{12}^2 c_{13}^2 e^{i\alpha} m_2
      + s_{13}^2 e^{i\beta} m_3 
\end{equation}
Eq.(\ref{defmeff}) contains a priori seven unknowns: Three mass eigenstates, 
two angles and two phases. With the help of data from neutrino oscillation 
experiments, one can trade two mass eigenstates for the observed 
$\Delta m^2_{\rm Atm}$ and $\Delta m^2_{\odot}$ and relate the two 
angles to the solar ($\theta_{\odot}$) and reactor angle ($\theta_{R}$). 
For the case of normal hierarchy, $m_{\nu_1} \le m_{\nu_2} \le m_{\nu_3}$, 
eq.(\ref{defmeff}) can then be written as
\begin{equation}\label{NormHier}
\meff =  c_{\odot}^2 c_{R}^2 m_{\nu_1} 
       + s_{\odot}^2 c_{R}^2 e^{i\alpha} 
         \sqrt{m_{\nu_1}^2+\Delta m^2_{\rm \odot}} 
       + s_{R}^2 e^{i\beta}  
         \sqrt{m_{\nu_1}^2+\Delta m^2_{\rm \odot}+\Delta m^2_{\rm Atm}},
\end{equation}
while for the case of inverse hierarchy, $m_{\nu_3} \le m_{\nu_1} \le 
m_{\nu_2}$, it is given by 
\begin{equation}\label{InvHier}
\meff =  c_{\odot}^2 c_{R}^2 
         \sqrt{m_{\nu_3}^2-\Delta m^2_{\rm \odot}+\Delta m^2_{\rm Atm}}
       + s_{\odot}^2 c_{R}^2 e^{i\alpha} 
         \sqrt{m_{\nu_3}^2+\Delta m^2_{\rm Atm}} 
       + s_{R}^2 e^{i\beta} m_{\nu_3}  
\end{equation}

\begin{figure}[h]
\vskip-5mm
\includegraphics[width=18pc]{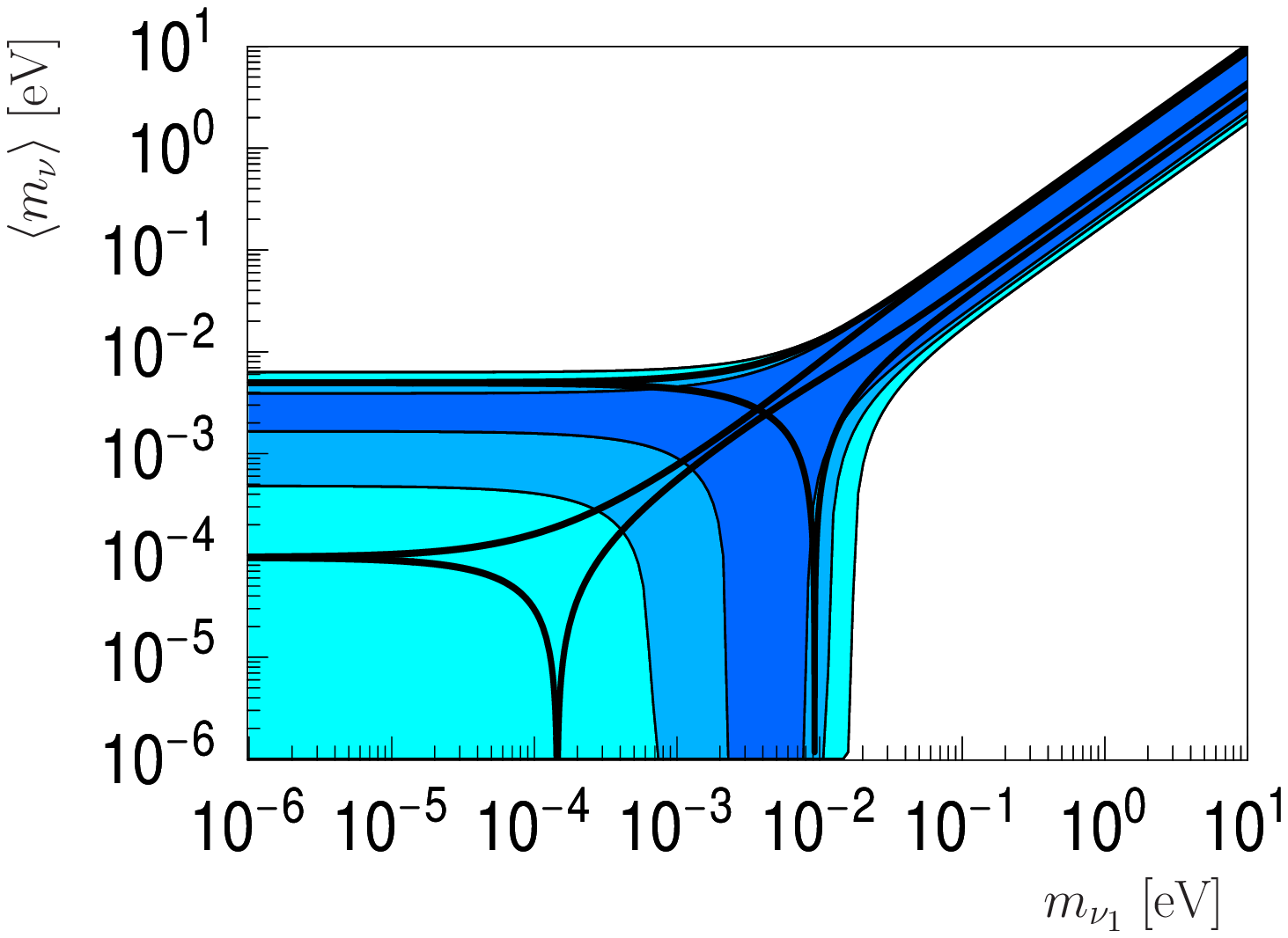}\hspace{2pc}%
\includegraphics[width=18pc]{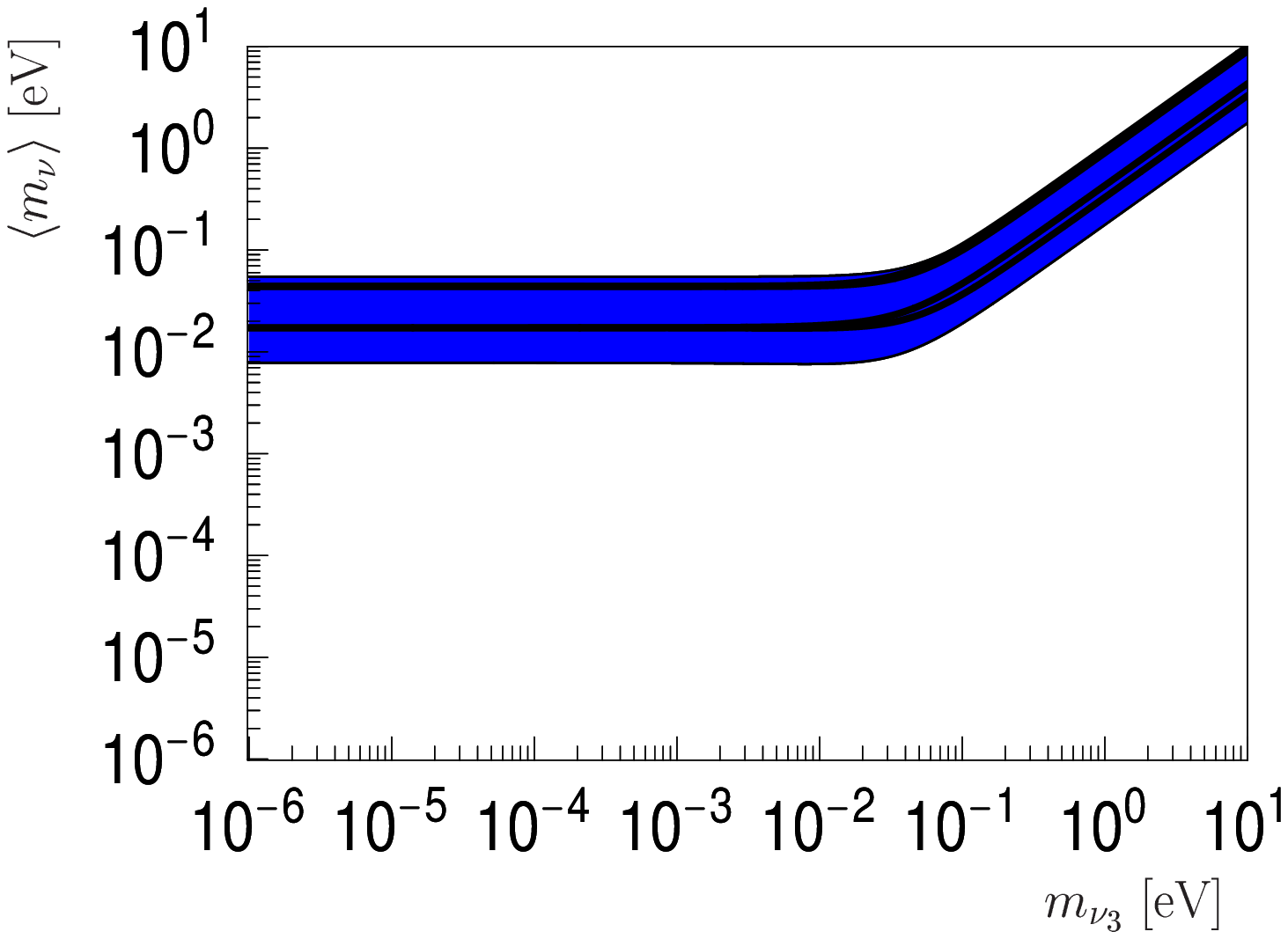}
\caption{\label{fig:NandI}{Allowed range of $\meff$ as a function of 
the lightest neutrino mass eigenvalue. To the left normal hierarchy, 
to the right inverse hierarchy. To calculate the allowed range of 
$\meff$ the 3 $\sigma$ c.l. intervals on the oscillation parameters have 
been used \cite{Maltoni:2004ei}, except for the case of normal hierarchy, 
for which 3 different cases for the upper limit on $s_R^2$ are shown. 
These are $s_R^2 \le 0.04$ (light blue), $s_R^2 \le 0.025$ (medium blue), 
$s_R^2 \le 0.005$ (darker blue).}}
\end{figure}

\begin{figure}[h]
\includegraphics[width=22pc]{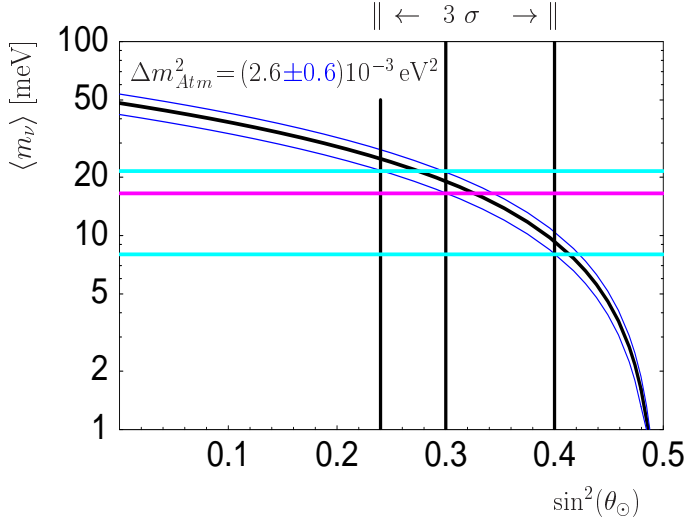}\hspace{2pc}%
\begin{minipage}[b]{14pc}
\caption{\label{fig:invlim}Lower Limit on $\meff$ in the case 
of inverse hierarchy as a function of the solar mixing angle 
$\sin^2\theta_{\odot}$ for three different values of $\Delta m^2_{\rm Atm}$, 
i.e. best fit point $\pm$ 3 $\sigma$ allowed range. The vertical black 
lines indicate the current best fit point and the 3 $\sigma$ c.l. allowed 
range of $s_{\odot}^2\equiv \sin^2(\theta_{\odot})$. The worst case, i.e. 
the most conservative limit, is found for $\sin^2\theta_{\odot}^{\rm Max}$ 
and $(\Delta m^2_{\rm Atm})^{\rm Min}$, currently $\meff \ge 8$ meV.
}
\end{minipage}
\end{figure}

Fig. (\ref{fig:NandI}) shows the resulting allowed range of $\meff$ 
for both, normal and inverse hierarchy, taking into account the latest 
results from a global fit to all neutrino oscillation data 
\cite{Maltoni:2004ei}. The lower limit on $\meff$, which appears in the 
case of inverse hierarchy, can be understood trivially. For $m_{\nu_3} =0$ 
and $\alpha = \pi$ eq. (\ref{InvHier}) reads approximately
\begin{equation}\label{InvLim}
\meff \simeq c_R^2(c_{\odot}^2-s_{\odot}^2) \sqrt{\Delta m_{Atm}^2}.
 \end{equation}
Thus, as soon as data tells us that $s_{\odot}^2< \frac{1}{2}$, exact 
cancellation is no longer a possibility. This statement remains true 
for any finite $m_{\nu_3}$, simply because $s_R^2 < \cos(2\theta_{\odot})$ 
is guaranteed by data nowadays. Fig. (\ref{fig:invlim}) shows how this 
lower limit evolves with future data from neutrino oscillation experiments. 
A possible future smaller upper bound on $s^2_{\odot}$ would 
make it easier for $\znbb$ decay experiments to rule out inverse hierarchy. 

There is no such simple quantitative lower limit for the case of normal 
hierarchy. Fig. (\ref{fig:NandI}), to the left, aims at demonstrating this 
point. If $m_{\nu_1}\equiv 0$, a lower limit appears if 
\begin{equation}\label{NrmLim}
s_R^2 \le \frac{\sqrt{\Delta m^2_{\odot}} s_{\odot}^2}
          {\sqrt{\Delta m^2_{\odot}+\Delta m^2_{\rm Atm}} +
           \sqrt{\Delta m^2_{\odot}} s_{\odot}^2} \sim 0.034
 \end{equation}
However, from this superficial look at the data at the point 
$m_{\nu_1} = \tan^2\theta_{\odot} m_{\nu_2}$ {\rm exact} cancellation 
yielding $\meff \equiv 0$ seems possible. However, this is equivalent 
to saying $M_{ee}^{\nu} \equiv 0$ and it is exactly this possibility  
which is ruled out by the ``extended black box'' theorem \cite{hir2006}.

\begin{figure}[h]
\vskip-5mm
\includegraphics[width=22pc]{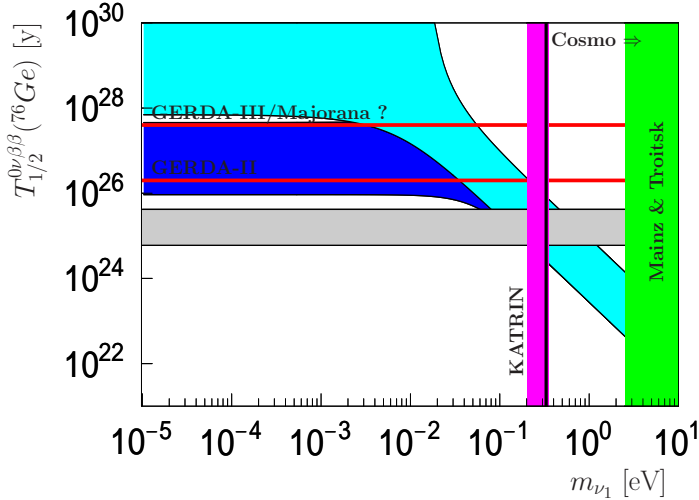}\hspace{2pc}%
\begin{minipage}[b]{14pc}
\caption{\label{fig:exp}Summary of experimental data on the absolute 
neutrino mass scale and the half-life of $^{76}$Ge $\znbb$ decay. 
For discussion see text.}
\end{minipage}
\end{figure}

In fig. (\ref{fig:exp}) finally a summary of the current status of 
various experimental attempts on measuring/limiting the absolute scale 
of neutrino masses is given. The light and darker blue areas are allowed 
for the $\znbb$ decay half live of $^{76}$Ge for normal and inverse 
hierarchy, calculated with matrix elements from \cite{Muto:1997qx}. Note, 
that matrix elements from \cite{Rodin:2006yk} lead to slightly larger 
half-lives, see also the discussion in \cite{simkovic}.
The green area labeled ``Mainz \& Troitsk'' shows the latest upper limits 
derived from endpoint measurements in $^{3}$H decay 
\cite{Kraus:2004zw,Lobashev:2001uu}. The bar labeled ``KATRIN'' represents 
the expected sensitiviy of the next generation $^{3}$H experiment KATRIN 
\cite{Osipowicz:2001sq}. Note, that KATRIN claims a final sensitivity of 
$m_{\nu_e} \sim 0.2$ eV ($@$ 90 \% c.l.) or a 5 $\sigma$ discovery threshold 
of $m_{\nu_e} \sim 0.35$ eV. Various limits on the absolute neutrino mass 
scale from cosmology have been published recently, derived from 
CMB data combined with information from large scale structure surveys. 
For three generations of neutrinos numbers ranging from 
$\sum_i m_{\nu_i} \sim 0.4-2.0$ eV, depending on input and bias, 
have been published. For a detailed discussion see, for example, the 
review \cite{Lesgourgues:2006nd}. The horizontal gray band indicates the 
range of the finite $T_{1/2}^{\znbb}$ claimed by some members of the 
Heidelberg-Moscow experiment \cite{Klapdor-Kleingrothaus:2004ge}. Note that 
this result is highly controversial, see for example the discussion  
by Barabash in \cite{bbexp}. The vertical red lines indicate 
the sensitivity of two future Ge experiments. GERDA \cite{Abt:2004yk} is 
currently in phase I, phase II is funded. In the future Majorana 
\cite{Gaitskell:2003zr} and/or GERDA phase III can test the range 
allowed by inverse hierarchy. 

\section{Conclusions}

Lower limits on the $\znbb$ decay half live can be used to constrain various 
particle physics parameters. However, from the point of view of particle 
physics it would be interesting to determine the {\em dominant} contribution 
to $\znbb$ decay. Very little work has been done in this direction. Angular 
correlations between the eletrons \cite{Doi:1985dx} or a comparative study 
of $0\nu\beta^-\beta^-$ and $0\nu\beta^+/EC$ decay \cite{Hirsch:1994es} 
might be able to disentangle left-left and left-right-handed combinations 
of currents (of the long range type). However, other contributions to 
$\znbb$ decay possibly exist and ultimately it might be that only a 
combination of various different pieces of experimental data will provide 
the correct and final answer.

\medskip
{\bf Acknowledgments}

I would like to thank S.G. Kovalenko and J.W.F. Valle for various 
discussions on the subject. Financial support by Spanish grant 
FPA2005-01269, by European Commission Human Potential Program RTN 
network MRTN-CT-2004-503369 and the EU Network of Astroparticle Physics
(ENTApP) WP1, as well as the spanish MCyT Ramon y Cajal 
program is acknowledged.

\end{document}